\documentclass[prl, notitlepage, superscriptaddress,  reprint]{revtex4-1}
\usepackage{amssymb}
\usepackage{amsmath}
\usepackage{braket}
\usepackage{graphicx}
\usepackage[usenames,dvipsnames]{color}
\usepackage{tensor}
\usepackage{empheq}
\usepackage{capt-of}
\usepackage[normalem]{ulem} % either use this (simple) or
\usepackage{soul} % 

\definecolor{myorange}{RGB}{199.24, 87.48, 47.80}
\usepackage[colorlinks,bookmarks=false,citecolor=NavyBlue,linkcolor=OliveGreen,urlcolor=blue]{hyperref}
\newcommand{\be}{\begin{equation}}
	\newcommand{\ee}{\end{equation}}
\newcommand{\ba}{\begin{aligned}}
	\newcommand{\ea}{\end{aligned}}
\newcommand{\bea}{\begin{eqnarray}}
\newcommand{\eea}{\end{eqnarray}}

\newcommand{\bw}{\begin{widetext}}
	\newcommand{\ew}{\end{widetext}}

\newcommand{\xt}{\zeta}

\begin{document}
%%%%%%%%%%%%%%%%%%%%%%%%%%%%%%%%%%%%%%%%%%%%%%%%%%%%%%%%%%%%%%%%%%%%%%%%%%%%%%
\title{Universal broadening of the light cone in low-temperature transport}
%%%%%%%%%%%%%%%%%%%%%%%%%%%%%%%%%%%%%%%%%%%%%%%%%%%%%%%%%%%%%%%%%%%%%%%%%%%%%%

\author{Bruno Bertini}
\author{Lorenzo Piroli}
\author{Pasquale Calabrese}
\affiliation{SISSA and INFN, via Bonomea 265, 34136 Trieste, Italy. }

\begin{abstract}
We consider the low-temperature transport properties of critical one-dimensional systems which can be described, at equilibrium, by a Luttinger liquid. We focus on the prototypical setting where two semi-infinite chains are prepared in two thermal states at small but different temperatures and suddenly joined together. At large distances $x$ and times $t$, conformal field theory characterizes the energy transport in terms of a single light cone spreading at the sound velocity $v$. Energy density and current take different constant values inside the light cone, on its left, and on its right, resulting in a three-step form of the corresponding profiles as a function of $\zeta=x/t$. 
Here, using a non-linear Luttinger liquid description, we show that for generic observables this picture is spoiled as soon as a non-linearity in the spectrum is present. In correspondence of the transition points $x/t=\pm v$ a novel universal region emerges at infinite times, whose width is proportional to the temperatures on the two sides. In this region, expectation values have a different temperature dependence and show smooth peaks as a function of $\zeta$. We explicitly compute the universal function describing such peaks. In the specific case of interacting integrable models, our predictions are analytically recovered by the generalized hydrodynamic approach.
\end{abstract}

\maketitle

The existence of universal phenomena is arguably the most fascinating aspect of many-body physics. As a genuinely collective behavior, universality cannot be understood based uniquely on the knowledge of elementary constituents and represents, as such, a fundamental fact of nature. In the quantum realm, universal effects emerge in the low-energy description of critical systems as adequately captured, in two and three spatial dimensions, by the Landau's theory of Fermi liquids~\cite{nozieres-99}. The situation is different in one dimension, where interactions play a special role, resulting in a strong collective nature of elementary excitations, and generating a plethora of fascinating new effects \cite{giamarchi-04}. In this case, the low-energy description is provided by the universal theory of Luttinger liquids \cite{Hald81, GNTbook, giamarchi-04}, now routinely applied in the study of several interacting systems.
Under the basic assumption that, at low energies, the spectrum of microscopic excitations can be linearized, the relevant physics (in a renormalization group sense) of the interacting many-body system is captured by an {\it emergent free} theory, representing the collective modes of the system~\cite{giamarchi-04,Hald81, GNTbook}. Besides providing a conceptually unifying point of view, the existence of universal effects is also of practical importance:
first-principle calculations in interacting many-body systems are overwhelmingly hard, and the presence of universality allows one to provide precise quantitative predictions based on a few model-dependent phenomenological parameters.

While the power of universality has been traditionally exploited at equilibrium, recent research has been increasingly interested in its emergence out of equilibrium. For example, a universal description has been found to apply for the spreading of 
entanglement~\cite{CC, CC-review,ac-17a} and correlations~\cite{CC-review,cc-06} after a quantum quench. Here we focus on another significant example in the context of nonlinear transport in one dimension: we consider two semi-infinite systems prepared at different temperatures, $T_L$ and $T_R$, and study the dynamics after they have been suddenly joined together at time $t=0$~\cite{BeDo16Review, VMreview}. %In this setting, at large distances $x$ {\color{red} from the junction} and times $t$, generic observables describe nontrivial {\it profiles} as a function of the ratio~$\zeta = x/t$.
For $t>0$ the two halves start to mix originating a non-equilibrium region spreading out from the junction. This region is known as \emph{light cone} and when a quasiparticle description applies is tipically of size $\sim t$. 
In this case, at large times $t$, generic local observables become quasi-stationary and their {\it profiles} become functions of the ratio~$\zeta = x/t$ where $x$ is the distance from the junction.

\begin{figure}
	\includegraphics[scale=0.89]{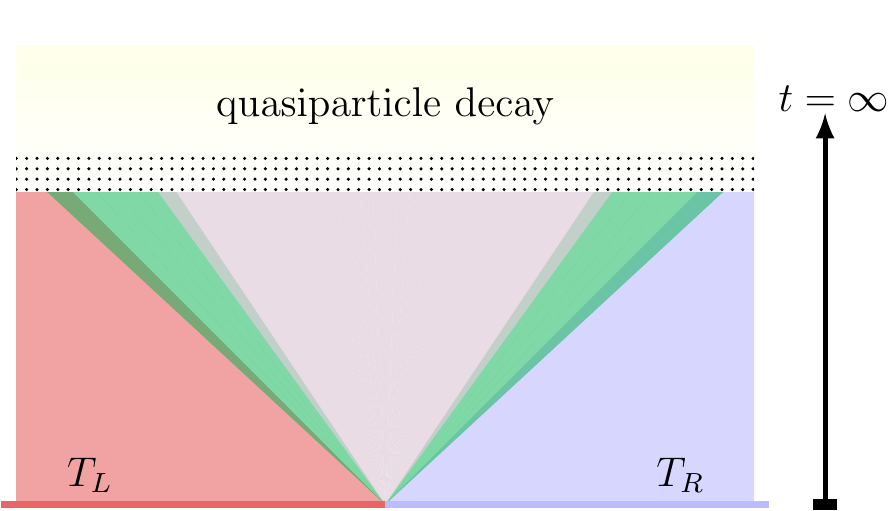}
	\caption{Pictorial representation of low-temperature transport dynamics in a critical system. After a rapid transient time following the junction, the dynamics is accurately described by a nonlinear Luttinger liquid. %As discussed in the main text, 
Profiles of local observables display a three-step form as a function of $\zeta=x/t$, corresponding to a %single 
light cone propagating at velocity $v$. 
At its edges a novel universal region emerges characterized by smooth peaks (green areas in the figure). They are genuine nonlinear effects and their width is proportional to the temperature. 
After a typical time scale $\tau_e$, other irrelevant terms become important, spoiling the quasiparticle 
picture and starting an eventual diffusive dynamics~\cite{BeDo16Review}}
	\label{fig:sketch}
\end{figure}

For a large intermediate time window (see Fig.~\ref{fig:sketch}), the low temperature dynamics of gapless systems 
is controlled by the equilibrium low energy modes and it is then universal. A conformal field theory (CFT) description for such situations~\cite{BeDo16Review,BeDo12, BeDo15,BDLS15} provides
quantitative predictions for the transport of energy: the profiles of its density and current display a three-step form, with sharp transitions in correspondence of the sound velocity. This is consistent with the fact that a linear spectrum can only give rise to one velocity of propagation, resulting in a single sharp light cone. Moreover, once one subtracts the ground-state values, the profiles are fully determined by a few quantities: the two temperatures, the sound velocity, and the central charge of the underlying CFT. Such results have been verified both in numerical \cite{KaIM13,DVMR14} and analytical studies, in free~\cite{DVBD, PeGa17, Korm17} and interacting~\cite{CaDY16, inpreparation, inpreparation2} integrable models. The latter could be investigated by the generalized hydrodynamic theory recently introduced in \cite{CaDY16, BCDF16} (see also \cite{DoYo17,DDKY17,IlDe17,Alba17,PDCB17,BVKM17,Bulchandani-17,Fago17-charges,BaDe17,Fago17-higher} for interesting developments).

The predictions of Refs.~\cite{BeDo12, BeDo15} are limited to the transport of energy and cannot be applied to other quantities such as, {\it e.g.}, magnetization or particle density. In this letter we provide {\it universal} predictions for generic local observables. Using a nonlinear Luttinger liquid description~\cite{ImGl09-science,ImGl09,ImScGl-review, KaPS15}, we show that a {\it novel universal region} emerges, where profiles correspond to a universal function of $\zeta$ peaked around the sound velocity. We discuss the range of applicability of our predictions for critical systems, and test them in the interacting integrable case.

\paragraph{Low-energy Luttinger liquid description.} 
We consider a gapless system in the Luttinger universality class. Assuming that, at small temperatures, only low energy modes contribute to the dynamics, we can compute the profiles of local observables using a linear Luttinger liquid approach. To this end, we adopt a description in terms of free fermionic (rather than bosonic) particles with linear dispersion relation. This point of view, which is often called ``refermionization", is well suited to include nonlinear effects~\cite{Rozhkov, ImScGl-review}. Two species of free quasiparticles are present, the left and right movers, with energy ${\varepsilon_{r}(k)=v|k|}$ and quasi-momentum ${p_r(k)=r|k|}$. Here the phenomenological parameter $v$ plays the role of a sound velocity, while the two signs ${r=+}$ and ${r=-}$ correspond to right and left movers respectively. Note that the particles velocity is given by ${v_{r}(k)=\partial\varepsilon_{r}(k)/\partial p_{r}(k)=r v}$; quasi-particles with $rk < 0$ are interpreted as ``hole'' excitations for the system.

Consider a reference frame moving at velocity $\zeta>0$ away from the origin. If $\zeta>v_r(k)$ particles of species $r$ coming from the left half of the system never reach our reference frame and the density $n_{r,\zeta}(k)$ is equal to the one of the right half, $n_{r}^R(k)$. By a similar reasoning, one can conclude that in the case $\zeta<v_r(k)$ the density $n_{r,\zeta}(k)$ coincides with the one of the left half, $n_{r}^L(k)$. Repeating this argument for $\zeta<0$ we obtain
\bea
n_{r,\zeta}(k)=(n_{r}^L(k)-n_{r}^R(k))\Theta(v_r(k)-\zeta)+n_{r}^R(k)\,.
\label{eq:n_zeta}
\eea
Here $\Theta(k)$ is the Heaviside theta function and for the specific problem under consideration the distributions $n_r^{L/R}(k)$ are given by the thermal Fermi distributions $n^{L/R}_r(k)=(1+e^{\varepsilon_r(k)/T_{L/R}})^{-1}$. 

At large times, local observables at distance $x$ from the junction are characterized by a locally quasi-stationary state~\cite{BF-defect}, which is entirely specified by the corresponding densities of quasiparticles $n_{r,\zeta}(k)$  \eqref{eq:n_zeta}, with $\zeta=x/t$.
In the simplest example of the energy density $\langle {\bf e}\rangle_{\zeta}$, we have
\be
\delta\langle {\bf e}\rangle_{\zeta} =\sum_{r=\pm}\int_{-\infty}^{\infty}\frac{dk}{\rm 2 \pi}\,n_{r,\zeta}(k)\varepsilon_{r}(k)\,, 
\label{eq:profiles_observables}
\ee
where we denoted by $\delta \langle {\bf e}\rangle_{\zeta}$ the difference between $\langle {\bf e}\rangle_{\zeta}$ and the corresponding ground-state expectation value. The integral  \eqref{eq:profiles_observables} can be computed exactly and a simple structure for the profiles emerges
\bea
\delta\langle {\bf e}\rangle_{\zeta}=&&\frac{\pi T^2_L}{6v}\Theta(-\zeta-v)
+\frac{\pi T^2_R}{6v}\Theta(\zeta-v)
\nonumber\\
&&+\frac{\pi (T^2_L+T^2_R)}{12 v}\Theta(v-|\zeta|)\,.
\label{eq:LL_energyresult}
\eea
As expected, the conformal prediction of Refs.~\cite{BeDo12, BeDo15, BeDo16Review} is exactly recovered by the Luttinger liquid theory, characterized by a central charge $c=1$. Specifically, the profile displays a three-step form, corresponding to a single light cone propagating at velocity $v$. 

A similar calculation can be performed for an arbitrary local conserved density or current ${\bf q}$. The expectation value of such an operator is always written as in \eqref{eq:profiles_observables}, where $\varepsilon_\pm(k)$ are replaced by some functions ${f_\pm(k)}$. To account for the fact that excitations with $rk<0$ are holes, these functions must satisfy $f_\pm(k)=\text{sgn}(k)g_\pm(k)$, where $g_\pm(k)$ are smooth functions of $k$. While the integral cannot in general be evaluated analytically, its leading behavior in the temperatures $T_{L/R}$ can be easily determined by expanding $f_\pm(k)$ on the left and on the right of $k=0$. In particular, one finds    
\bea
\delta \langle {\bf q}\rangle_{\zeta}=&&\frac{\pi(a+b)}{12v^2}T_L^2\Theta(-\zeta-v)+\frac{\pi(a+b)}{12v^2}T_R^2\Theta(\zeta-v)
\nonumber\\
&&+\frac{\pi (a T^2_L+ b T^2_R)}{12v^2}\Theta(v-|\zeta|)+O(T^4_{L/R})\,,
\label{eq:LL_result}
\eea
where $a=g_+^{\prime}(0)$, $b=g_-^{\prime}(0)$, and we denoted by $\delta \langle {\bf q}\rangle_{\zeta}$ the difference between $\langle {\bf q}\rangle_{\zeta}$ and the corresponding ground-state expectation value. From \eqref{eq:LL_result} we see that in the linear Luttinger liquid approximation the profiles of all the local conserved charges assume the same three-step form, determined uniquely by the sound velocity $v$. Moreover, the leading behavior of the individual plateaus is always $O(T_{L/R}^2)$, displaying observable-dependent amplitudes $a$ and $b$. 

\begin{figure}
	\includegraphics[scale=0.75]{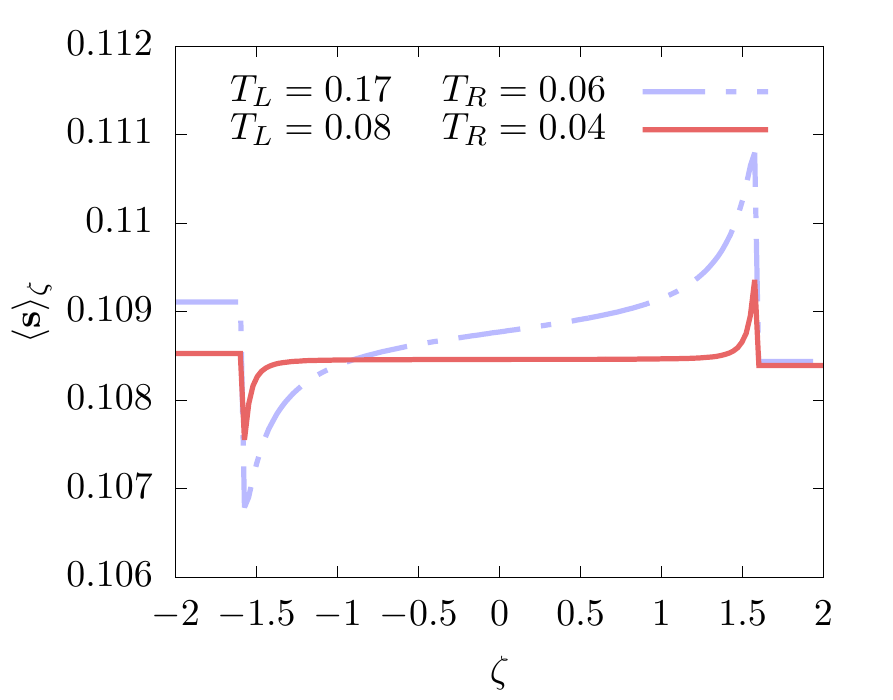}
	\caption{Low-temperature profiles in $\zeta=x/t$ of magnetization density ${\boldsymbol s_j= (1/2) \boldsymbol\sigma_{j}^{z}}$ in the gapless phase of the XXZ spin-1/2 chain. The Hamiltonian is ${\boldsymbol H= \frac{J}{4}\sum_{j=1}^{L}\left[\boldsymbol\sigma_{j}^{x}\boldsymbol\sigma_{j+1}^{x}+\boldsymbol\sigma_{j}^{y}\boldsymbol\sigma_{j+1}^{y}+\Delta \boldsymbol\sigma_{j}^{z}\boldsymbol\sigma_{j+1}^{z}\right]-h \sum_{j=1}^{L} \boldsymbol \sigma_j^z}$, where $\boldsymbol\sigma_{j}^{\alpha}$ are Pauli matrices. The plot corresponds to $\Delta=3$, $h=1.2$. As the temperature is lowered, the profiles are seen to approach the nonlinear Luttinger liquid prediction: a plateau emerges in the light cone, at the edges of which two peaks are clearly visible. For smaller temperatures, a quantitative comparison with our predictions is displayed in Fig.~\ref{fig:comparison}.}
	\label{fig:fullprofile}
\end{figure}

\paragraph{Nonlinear Luttinger liquids.} 
It is well established that a linear Luttinger liquid approximates the dynamics of a real critical Hamiltonian up to contributions that are irrelevant in the renormalization group sense~\cite{giamarchi-04, Hald81}. While this guarantees that such description gives the most relevant contribution to the large distance behavior of correlation functions on the ground state, one could wonder whether or not the irrelevant terms affect the characterization of low temperature transport. Our strategy here is to consider the effect of the next more relevant terms and self-consistently determine which of the above predictions are robust. %This will give a strong support to their universality.

Taking into account additional irrelevant terms modifies the fermionic quasi-particle Hamiltonian in two ways~\cite{ImGl09-science,ImGl09,ImScGl-review,Rozhkov}.
First, interaction terms between the quasi-particles emerge. In the low temperature regime, however, the density of excitations decreases as the temperatures are lowered, so that these terms are expected to be perturbatively small and can be neglected at the leading order~\cite{note1}. Second, the dispersion relation of fermionic quasi-particles is deformed by acquiring a curvature; namely for small $k$
\be
\varepsilon_r(k)= v |k|+\frac{r}{2m_{\ast}}|k| k +O(k^3)\,,\quad p_r(k)=r|k|\,,
\label{eq:quadratic_dispersion}
\ee
where a {\it single} additional phenomenological parameter $m_{\ast}$ is introduced, and where we still assume $\lim_{k\to\pm \infty }\varepsilon_r(k)=+\infty$. Importantly, this effect is not perturbatively small in the temperatures.

The nonlinearity in the dispersions gives a linear $k$-dependence to the velocities $v_r(k)$, modifying the light-cone  structure of the profiles. As opposed to the linear case, given $\zeta>0$ ($\zeta<0$) there are always quasiparticles coming from the left (right) reservoir with ${v_{r}(k)>\zeta}$ (${v_r(k)<\zeta}$). However, if ${||\zeta|-v|\gg T_{L/R}/(m_\ast v)}$ this does not affect much the final result, because such quasiparticles have exponentially vanishing weight. In this regime, at the leading order, the profiles are still expressed by the function \eqref{eq:LL_result}, but the prefactors have to be modified as ${a=g^{\prime}_+(0)-(m_\ast v)^{-1} g^{\phantom{\prime}}_+(0)}$ and ${b=g_-^{\prime}(0)+(m_\ast v)^{-1} g^{\phantom{\prime}}_-(0)}$. This shows that outside the region $||\zeta|-v|\sim T_{L/R}/(m_\ast v)$ the functional dependence of the leading order \eqref{eq:LL_result} is universal, while the prefactors $a$ and $b$ are not, as they are modified by the curvature $m_{\ast}$. Note that for the energy density and current $\varepsilon_\pm(0)=0$ implies $g_\pm(0)=0$~\cite{supmat}, confirming that the conformal prediction is robust also for the numerical values of the three plateaux.
\begin{figure*}[t!]
	\begin{tabular}{lll}
		\hspace{-0.25cm}\includegraphics[width=0.34\textwidth]{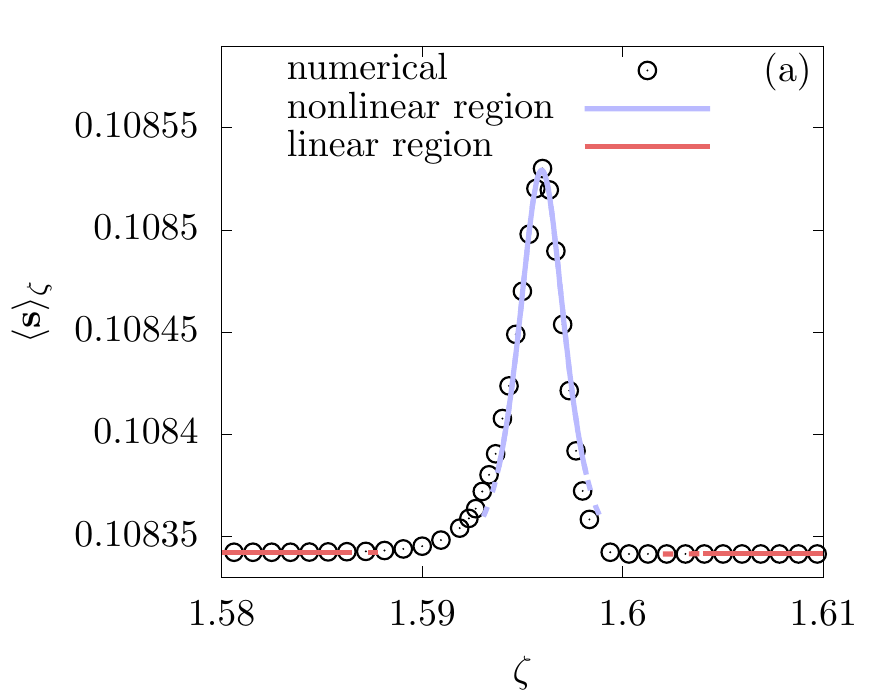} & \hspace{-0.5cm}\includegraphics[width=0.34\textwidth]{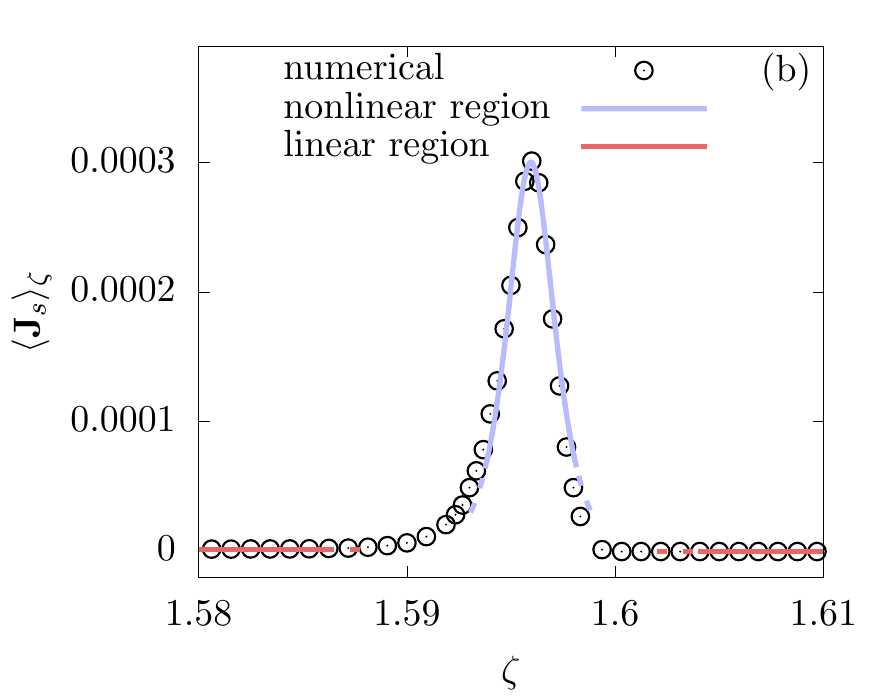}& \hspace{-0.5cm}\includegraphics[width=0.34\textwidth]{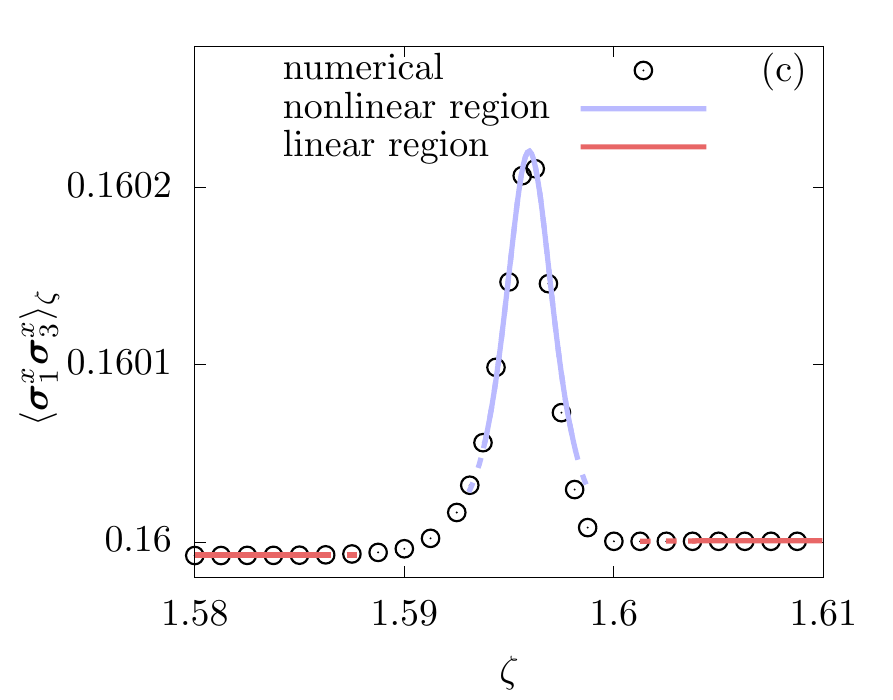}\\
	\end{tabular}
	\caption{Low-temperature profiles in $\zeta=x/t$ of magnetization density, magnetization current, and a local correlator in the gapless phase of the XXZ spin-1/2 chain, with the same Hamiltonian parameters as in Fig.~\ref{fig:fullprofile}.  The two temperatures of the two halves are $(T_L,T_R)=(1/125,1/250)$. The figure shows the comparison between the exact result from explicit numerical solution of the continuity equations~\eqref{eq:continuity} (circles)  and the predictions \eqref{eq:LL_result} (red solid line) and \eqref{eq:transitionregion} (blue solid line). For subfigures (a) and (b), the multiplicative factors $a,b$ and $d$ are computed exactly via a low temperature expansion~\cite{inpreparation}, while for subfigure (c) are obtained by fitting. At the transition between the two regions described by \eqref{eq:LL_result} and \eqref{eq:transitionregion} the observables display a non-universal behavior%, of higher order in $T$
~\cite{inpreparation}.}
	\label{fig:comparison}
\end{figure*}
Something much more interesting happens in the regime ${||\zeta|-v|\sim T_{L/R}/(m_\ast v)}$. 
Here the nonlinearity in the dispersion has a nonperturbative effect in the structure of the profiles, originating a qualitatively different 
low-temperature behavior, {\it cf}. Fig.~\ref{fig:fullprofile}. Specifically, by computing the leading order in the temperature of $\delta\langle {\bf q}\rangle_\zeta$ for $\zeta-v\sim T_{L/R}/(m_\ast v)$, we find~\cite{supmat} 
\be
\delta\langle {\bf q} \rangle_\zeta=\frac{ d\, T_L}{2\pi v^2} {\cal D}_{\frac{T_R}{T_L}} \left[\frac{m_{\ast}v}{T_L}(\xt-v)\right]+O(T_{L/R}^2)\,,
\label{eq:transitionregion}
\ee
where
\be
{\mathcal D}_\eta (z)\equiv  \log(1+e^{z})-\eta\log(1+e^{z/\eta})\,,
\label{Dz}
\ee
and $d={\rm sgn} (m_\ast )g_+(0)$. An analogous expression is found for $\zeta+v\sim T_{L/R}/(m_\ast v)$.

Equation~\eqref{eq:transitionregion} implies that the profiles of charge densities and currents are all proportional to the smooth function $\mathcal{D}_\eta(z)$, which is independent of all phenomenological parameters. According to the nonlinear Luttinger liquid theory~\cite{ImScGl-review}, we expect the functional form of \eqref{eq:transitionregion} to be universal: all the less relevant terms we neglected give only higher order corrections in $T_{L/R}$.

The function $\mathcal{D}_\eta(z)$ displays a peaked form as a function of $z$ with width $\Delta z\sim1$. Therefore, generic profiles deviate from those found in CFT due to the appearance of peaks of width $\sim{T_{L/R}}/{m_{\ast}v}$ at the edges of the lightcone, \emph{cf.} Figs.~\ref{fig:fullprofile} and \ref{fig:comparison}. We note that, contrary to \eqref{eq:LL_result}, the term \eqref{eq:transitionregion} is linear in $T_{L/R}$ rather than quadratic and is then the most relevant feature at low temperatures. Such term vanishes for energy density and current, once again in agreement with the conformal predictions~\cite{BeDo12}.

\paragraph{Generic local observables.}

Our results demonstrate that in the regions ${||\zeta|-v|\gg T_{L/R}/(m_\ast v)}$ and ${||\zeta|-v|\sim T_{L/R}/(m_\ast v)}$ all local conserved densities $\{\boldsymbol{ q}_j\}$ are proportional to the same functions. Based on this observation, we argue that all local observables display the same behavior. Indeed, on a given ray $\zeta$ the expectation value of a local operator $\boldsymbol{\mathcal O}$ can be expressed in terms of a ray-dependent generalized Gibbs ensemble~\cite{BF-defect,Rigol07,ViRi16,EsFa16}, where the Lagrange multipliers are fixed by the expectation values of the conserved densities. Inverting the mapping between Lagrange multipliers and densities~\cite{CaDY16} we can write $\braket{\boldsymbol{\mathcal O}}_\zeta=\mathcal{F}(\{\braket{\boldsymbol{ q}_j}_\zeta\})$, where $\mathcal{F}$ depends on $\boldsymbol{\mathcal O}$. At low temperatures we have $\braket{\boldsymbol{ q}_j}_\zeta=\braket{\boldsymbol{ q}_j}_{\rm GS}+\delta\braket{\boldsymbol{ q}_j}_\zeta$, where $\braket{\boldsymbol{ q}_j}_{\rm GS}$  is the ground state value of the density and $\delta\braket{\boldsymbol{ q}_j}_\zeta$ is given either by \eqref{eq:LL_result} or \eqref{eq:transitionregion} and is $\sim T_{L/R}^a$ with $a=1,2$ depending on the regime.  We can then formally expand the expectation value of $\boldsymbol{\mathcal O}$ in $T_{L/R}$
\be
\braket{\boldsymbol{\mathcal O}}_\zeta=\braket{\boldsymbol{\mathcal O}}_{\rm GS}+\sum_j \delta\braket{\boldsymbol{ q}_j}_\zeta \partial_{\braket{\boldsymbol q_j}} \mathcal{F}(\{\braket{\boldsymbol{ q}_j}_{\rm GS}\})\,,
\ee
where we neglected higher orders. As all $\delta\braket{\boldsymbol{ q}_j}_\zeta$ are proportional to the same function, so is $\braket{\boldsymbol{\mathcal O}}_\zeta-\braket{\boldsymbol{\mathcal O}}_{\rm GS}$.

\paragraph{Interacting systems: the integrable case.}

The above predictions can be tested in the case of interacting integrable models admitting a thermodynamic Bethe ansatz (TBA)~\cite{takahashi, korepin} description. Indeed, in such systems Eq.~\eqref{eq:n_zeta} has an exact generalization~\cite{CaDY16, BCDF16}, whose results can be compared to the nonlinear Luttinger liquid predictions. 

The generalization of \eqref{eq:n_zeta} is based on the idea that also interacting integrable systems have stable quasiparticle excitations, possibly of different species. The densities $n_r(k)$, measuring the number of quasiparticle excitations in free systems, ought to be replaced by an appropriate measure. In particular, for a fixed species $n$ of quasiparticle excitations we have ${n_r(k) \rightarrow \vartheta_{n}(\lambda)\equiv{\rho_n(\lambda)}/({ \rho_n(\lambda)+\rho^h_n(\lambda)})}$. Here we introduced the filling function $\vartheta_{n}(\lambda)$, the root density $\rho_n(\lambda)$, and the density of holes $\rho^h_n(\lambda)$ of the $n$-th species. The functions  $\rho_n(\lambda)$ and $\rho^h_n(\lambda)$ are defined as densities of occupied and unoccupied ``rapidities" $\lambda$, which are conveniently used to parametrize eigenstates in integrable systems~\cite{takahashi, korepin}. These rapidities also parametrize the dispersion relation of the stable quasiparticle excitations: an excitation of the $n$th species and rapidity $\lambda$ has energy $\varepsilon_n(\lambda)$ and momentum $k^d_n(\lambda)$. Generalizing the expression for $v_r(k)$ in \eqref{eq:n_zeta}, its velocity is 
\be
v_{n}(\lambda)=\frac{\partial \varepsilon_n(\lambda)}{\partial k^d_n(\lambda)}=\frac{\varepsilon'_n(\lambda)}{2\pi (\rho_n(\lambda)+\rho^h_n(\lambda))}\,,
\ee
where we used $k_n^{d\,\prime}(\lambda)=2\pi (\rho_n(\lambda)+\rho^h_n(\lambda))$ (see \emph{e.g.}~\cite{bonnes14}). Note that $v_n(\lambda)$ depends non-trivially on the state (it fulfills a non-trivial set of integral equations involving $\{\rho_n(\lambda)\}_{n=1,2,\ldots}$) and, accordingly, gets a $\xt$ dependence in the transport problem under exam. With the described replacements, Eq.~\eqref{eq:n_zeta} is rewritten as ~\cite{CaDY16, BCDF16}
\be
\vartheta_{n,\xt}(\lambda)= (\vartheta_n^L(\lambda)- \vartheta_n^R(\lambda))\Theta(v_{n,\xt}(\lambda)-\xt)+ \vartheta_n^R(\lambda)\,.
\label{eq:continuity}
\ee

This equation has to be solved numerically using iterative schemes. At small temperature, however, one can use it to find the low temperature expansion of the profiles of local observables analytically~\cite{inpreparation}. Performing the calculation in the gapless phase of the XXZ spin-1/2 chain, we find that only the gapless species (that with $n=1$) of excitations contributes to the leading order result in $T_{L/R}$ and the profile have indeed the form \eqref{eq:LL_result} and \eqref{eq:transitionregion}, where the effective mass is computed as
\be
(m^*)^{-1}= \frac{\partial^2 \varepsilon^0_1(\lambda)}{\partial k_1^{d\,0}(\lambda)^2}\bigg|_{k_1^{d\,0}(\lambda)=k_F} = \frac{v^{0\prime}_1(B)v^0_1(B)}{\varepsilon^{0\prime}_1(B)}\,.
\ee
Here the superscript $0$ is for properties of excitations above the ground state, while $B$ is the ``Fermi rapidity" such that $\varepsilon^{0}_1(B)=0$. In Fig.~\ref{fig:comparison} we report the comparison between the predictions \eqref{eq:LL_result} and \eqref{eq:transitionregion} and the exact profiles obtained by numerical solution of \eqref{eq:continuity} for some relevant cases. Note that the observable reported in Fig.~\ref{fig:comparison} (c) is not a charge density or current, confirming that our predictions are valid for generic local observables.

Similar calculations can be repeated straightforwardly for all interacting integrable models whose low-energy spectrum is 
captured by a single Luttinger liquid. Consequently, our predictions can be verified analytically in many physically relevant systems, such as
one-dimensional Bose gases. Furthermore, our results are naturally extended to systems described by multiple Luttinger liquids~\cite{inpreparation2}.

\paragraph{Conclusions.}
We studied the dynamics of gapless systems in the Luttinger liquid universality class ensuing after junction of two halves prepared at small but different temperatures. We showed that a non-linearity in the dispersion produces a novel universal region of width $\sim T_{L/R}$ at the edge of the light cone in the profiles of local observables: in this region, the latter are described by the same universal function~\eqref{Dz}. We expect nonlinear effects to also influence low-energy transport for other quench protocols, as \emph{e.g.} those considered in~\cite{BCDF16,CDV17,sm-13,viti}. The form of the profiles will be, however, generically different from~\eqref{Dz}. Finally, it would be interesting to study how nonlinearities affect the low-energy transport of gapless systems belonging to other universality classes, such as the Ising model.

\begin{acknowledgments}
We thank Fabian Essler for stimulating discussions. BB and PC acknowledge financial support by the ERC under  Starting Grant 279391 EDEQS. 
\end{acknowledgments}

\onecolumngrid
%\appendix

%%%%%%%%%% Merge with supplemental materials %%%%%%%%%%
%%%%%%%%%% Prefix a "S" to all equations, figures, tables and reset the counter %%%%%%%%%%
\newpage

\newcounter{equationSM}
\newcounter{figureSM}
\newcounter{tableSM}
\stepcounter{equationSM}
\setcounter{equation}{0}
\setcounter{figure}{0}
\setcounter{table}{0}
\makeatletter
\renewcommand{\theequation}{\textsc{sm}-\arabic{equation}}
\renewcommand{\thefigure}{\textsc{sm}-\arabic{figure}}
\renewcommand{\thetable}{\textsc{sm}-\arabic{table}}

\onecolumngrid

%%%%%%%%%%%%%%%%%%%%%%%%%%%%%
\begin{center}
{\large{\bf Supplemental Material for\\ ``Universal broadening of the light cone in low-temperature transport''}}
\end{center}
%%%%%%%%%%%%%%%%%%%%%%%%%%%%%
Here we collect a detailed derivation of the results presented in the main text. Specifically, in Section~\ref{sec:linear_luttinger} we derive Eq. (4) of the main text while in Section \ref{sec:nonlinear_luttinger} we derive Eq. (6). 

\section{Setting of the problem}
We consider two semi-infinite critical sub-systems joined together at the origin. The two subsystems are initially at temperatures $T_L$ and $T_R$ which we assume to be small. We focus on the large-time limit of the profiles of local densities and currents of conserved charges. In this limit the profiles become functions only of the ``ray" $\zeta=x/t$, where $t$ is the time and $x$ is the distance from the junction. We generically indicate them as $\langle {\bf q}\rangle_{\zeta}$. 

We assume that at the leading order in $T_{L/R}$ the system can be described by a Luttinger liquid which we represent as a gas of free fermionic quasiparticles of two different species~\cite{Rozhkov}. These quasiparticles  have quasi-momentum
\be
p_{r}(k)=r|k|\,,
\ee
while we will consider two different forms for their energy $\varepsilon_{r}(k)$: linear in $k$ in Section~\ref{sec:linear_luttinger} and quadratic in Section~\ref{sec:nonlinear_luttinger}. Excitations ${r k>0}$ correspond to ``particles", while those with ${r k <0}$ correspond to ``holes". The group velocity of such excitations is then given by 
\be
v_{r}(k)=\frac{\partial \varepsilon_{r}(k)}{\partial p_r(k)}\,.
\label{eq:velocity}
\ee
Under this assumption the generic profile $\langle {\bf q}\rangle_{\zeta}$ can be written as follows   
\be
\langle {\bf q}\rangle_{\zeta} =\langle {\bf q}\rangle_{\rm GS}+\sum_{r=\pm}\int_{-\infty}^{\infty}\frac{dk}{\rm 2 \pi}\,n_{r,\zeta}(k) f_{r}(k) \,, 
\label{eq:profile}
\ee
where $n_{r,\zeta}(k)$ is the density of quasiparticles at ray $\zeta$, given by   
\be
\label{eq:LQSS}
n_{r,\zeta}(k)=\Theta(\zeta-v_{r}(k))n^R_{r}(k)+\Theta(v_{r}(k)-\zeta)n^L_{r}(k)\,,
\ee
where 
\be
n_{r}^{L/R}(k)=\frac{1}{1+e^{\varepsilon_{r}(k)/T_{L/R}}}\,.
\label{eq:thermaloccnum}
\ee
The function $f_{r}(k)$ is the elementary charge carried by the quasiparticle specified by $r$ and $k$. To take into account that holes give a contribution which is opposite to that given by particles, it is convenient to parametrise $f_{r}(k)$ as follows  
\be
f_r(k)=
\begin{cases}
	\,\,\, g_r(k)\,, & k>0\,,\\
	-g_r(k)\,, & k<0\,,\\
\end{cases}
\label{eq:mathcal_f_plus}
\ee
where $g_r(k)$ is a smooth function of $k$. For example, for the energy density $g_r(k)$ reads as 
\be
g_r(k)=\text{sign}(k) \varepsilon_{r}(k)\,,
\ee
while for the energy current it reads as 
\be
g_r(k)=\text{sign}(k) v_{r}(k) \varepsilon_{r}(k)\,.
\ee

\section{Linear Luttinger liquid}\label{sec:linear_luttinger}

Let us begin by assuming that the Luttinger liquid under consideration is linear, \emph{i.e.}, we consider a dispersion relation of the form
\be
\varepsilon_{r}(k)=v |k|\,,
\ee
where the parameter $v$ is the sound velocity of the model. Using Eq.~\eqref{eq:velocity} it is then immediate to find $v_r(k)=r v$. Plugging these definitions in Eq.~\eqref{eq:LQSS} and Eq.~\eqref{eq:thermaloccnum} we find 
\be
\langle {\bf q}\rangle_{\zeta} =\langle {\bf q}\rangle_{\rm GS}+\sum_{r=\pm}\Theta(\zeta-r v)\int_{-\infty}^{\infty}\frac{dk}{\rm 2 \pi}\,n^R_{r}(k)f_{r}(k)+\sum_{r=\pm} \Theta(r v -\zeta)\int_{-\infty}^{\infty}\frac{dk}{\rm 2 \pi}\,n^L_{r}(k)f_{r}(k)\,.
\label{eq:profilesLLL}
\ee
So we just need to compute
\be
\int_{-\infty}^{\infty}\frac{dk}{\rm 2 \pi}\,n^{L/R}_{r}(k)f_{r}(k)=\int_0^{\infty}\frac{dk}{2\pi} n^{L/R}_{r}(k)\left(g_{r}(k)-g_{r}(-k)\right)\,,\qquad\qquad\qquad r=\pm\,.
\ee
Taylor expanding the smooth function $g_r(k)$ around 0 we have 
\be
\int_{-\infty}^{\infty}\frac{dk}{\rm 2 \pi}\,n^{L/R}_{r}(k)f_{r}(k)=\frac{\pi T^2_{L/R}}{12 v^2}g^{\prime}_r(0)+O(T_{L/R}^4)\,.
\ee
Substituting back in \eqref{eq:profilesLLL} we obtain Eq. (4) of the main text.

\section{Nonlinear Luttinger liquid}
\label{sec:nonlinear_luttinger}

In this section we consider the case of a non-linear Luttinger liquid \cite{Rozhkov,ImGl09}. As discussed in the main text, in the limit of low temperatures, one is still allowed to study transport problems by employing a description in terms of free fermionic excitations. In this case, we consider a dispersion relation with a non-vanishing curvature; namely, for small values of $k$
\be
\varepsilon_{r}(k)=v |k|+\frac{r}{2m_\ast }|k|k+O(k^3)\,,
\label{eq:nonlineardisp}
\ee
and where we still assume $\lim_{k\to\pm \infty}\varepsilon_r(k)=+\infty$.

Taking $\varepsilon_{r}(k)$ as in Eq.~\eqref{eq:nonlineardisp} yields the following form of the group velocity 
\be
v_{r}(k)=r v +\frac{1}{m_\ast } k\,.
\label{eq:nonlinearvel}
\ee
Plugging the explicit formulae \eqref{eq:nonlineardisp} and \eqref{eq:nonlinearvel} for the group velocity and energy of quasiparticle excitations in Eq.~\eqref{eq:profile} we find 
\be
\langle {\bf q}\rangle_{\zeta} =\langle {\bf q}\rangle_{\rm GS}+\sum_{r=\pm}\int_{-\infty}^{k^r_\zeta}\frac{dk}{\rm 2 \pi}\,n^R_{r}(k)f_{r}(k)+\sum_{r=\pm} \int_{k^r_\zeta}^{\infty}\frac{dk}{\rm 2 \pi}\,n^L_{r}(k)f_{r}(k)\,, 
\label{eq:nonlinearprofiles}
\ee
where we introduced 
\be
k^r_\zeta = m_\ast (\zeta-rv)\,,
\ee
and assumed ${m_\ast>0}$. Let us focus on the term 
\be
\int_{-\infty}^{k^r_\zeta}\frac{dk}{\rm 2 \pi}\,n^R_{r}(k)f_{r}(k)=T_R\int_{-\infty}^{k^r_\zeta/T_R}\frac{dk}{\rm 2 \pi}\,\frac{1}{1+e^{\varepsilon_r(k T_R)/T_R}} f_{r}(T_R k)
\ee
We see that if $|k^r_\zeta|=O(T_{R/L}^a)$ with $a<1$ then up to exponential corrections
\bea
\int_{-\infty}^{k^r_\zeta}\frac{dk}{\rm 2 \pi}\,n^R_{r}(k)f_{r}(k)&=&T_R \Theta(\zeta-rv)\int_{-\infty}^{\infty}\frac{dk}{\rm 2 \pi}\,\frac{1}{1+e^{\varepsilon_r(k T_R)/T_R}} f_{r}(T_R k)\nonumber\\
&=&\frac{\pi T^2_R}{12 v^2}\left[g^{\prime}_r(0)-\frac{rg_r(0)}{m_\ast v}\right] \Theta(\zeta-rv)+O(T^4_R)\,.
\eea
Analogously, if $|k^r_\zeta|=O(T_{R/L}^a)$ with $a<1$, we find   
\be
\int^{\infty}_{k^r_\zeta}\frac{dk}{\rm 2 \pi}\,n^L_{r}(k)f_{r}(k)=\frac{\pi T^2_L}{12 v^2}\left[g^{\prime}_r(0)-\frac{rg_r(0)}{m_\ast v}\right] \Theta(rv-\zeta)+O(T^4_L)\,.
\ee
So we see that if $|k^r_\zeta|=O(T_{R/L}^a)$ with $a<1$ the introduction of the curvature has no effect of the functional dependence of the profiles. 

On the other hand, the curvature gives new qualitative features $O(T_{R/L})$ close to the edges of the light cone, \emph{i.e.} when $k^+_\zeta\sim T_{R/L}$ or $k^-_\zeta\sim T_{R/L}$. Let us consider for example $k^+_\zeta\sim T_{R/L}$, from \eqref{eq:nonlinearprofiles} we get  
\bea
\langle {\bf q}\rangle_{\zeta} &=& \langle {\bf q}\rangle_{\rm GS} + \int_{k^+_\zeta}^{\infty} \frac{dk}{2\pi}{n}^L_+(k) f_+(k)+\int_{-\infty}^{k^+_\zeta} \frac{dk}{2\pi}{n}^R_+(k)f_+(k)+O(T_{R/L}^2)\nonumber\\
&=& \langle {\bf q}\rangle_{\rm GS} +\int_{k^+_\zeta}^{\infty} \frac{dk}{2\pi}{n}^L_+(k) g_+(k)+\int_{0}^{k^+_\zeta} \frac{dk}{2\pi}{n}^R_+(k)g_+(k)-\int_{-\infty}^{0} \frac{dk}{2\pi}{n}^R_+(k)g_+(k)+O(T_{R/L}^2)\,,
\eea
where we assumed $k^+_\zeta>0$. Computing each term up to order $O(T^2)$ we obtain 
\bea
\int_{k^+_\zeta}^{\infty} \frac{dk}{2\pi}{n}^L_+(k) g_+(k)&=&\frac{T_L g_+(0)}{2\pi v}\int^\infty_{v k^+_\zeta/T_L}\frac{dx}{1+e^{x}}+O(T^2_{L})\,, \nonumber\\
&=&\frac{T_Lg_+(0)}{2\pi v} \log\left[1+\exp(m_\ast v (v-\zeta)/T_L)\right]+O(T_L^2)\,,\\
\int_{0}^{k^+_\zeta} \frac{dk}{2\pi}{n}^R_+(k)g_+(k)&=&\frac{T_R g_+(0)}{2\pi v} \left( \log 2- \log\left[1+\exp(m_\ast v (v-\zeta)/T_R)\right]\right)+O(T_R^2)\\
\int_{-\infty}^{0} \frac{dk}{2\pi}{n}^R_+(k)g_+(k)&=&\frac{T_Rg_+(0)}{2\pi v}\log 2+O(T^2_{R})\,.
\eea
Putting everything together we find
\be
\langle {\bf q}\rangle_{\zeta} = \langle {\bf q}\rangle_{\rm GS} + \frac{T_L g_{+}(0)}{2 \pi v^2}\mathcal{D}_{T_R/T_L}\left(\frac{m_\ast v}{T_L}(\zeta-v)\right)+O(T_{L/R}^2)\,,
\label{eq:nonlinresm>0}
\ee
where
\be
\mathcal{D}_\eta(z)=\log(1+e^{z})-\eta \log(1+e^{z/\eta})\,.
\ee
The same result is found for $k^+_\zeta<0$. Repeating the calculation for $m_\ast<0$ we find  
\be
\langle {\bf q}\rangle_{\zeta} = \langle {\bf q}\rangle_{\rm GS} - \frac{T_L g_{+}(0)}{2 \pi v^2}\mathcal{D}_{T_R/T_L}\left(\frac{m_\ast v}{T_L}(\zeta-v)\right)+O(T_{L/R}^2)\,.
\label{eq:nonlinresm<0}
\ee
Equations \eqref{eq:nonlinresm>0} and \eqref{eq:nonlinresm<0} prove Eq. (6) of the main text. Finally, assuming instead $k^-_\zeta\sim T_{R/L}$ we find the contribution on the other edge of the light cone, putting all together we have 
\bea
\langle {\bf q}\rangle_{\zeta} &=& \langle {\bf q}\rangle_{\rm GS} +\text{sign}({m_\ast}) \frac{T_L g_{+}(0)}{2 \pi v^2}\mathcal{D}_{T_R/T_L}\left(\frac{m_\ast v}{T_L}(\zeta-v)\right)\nonumber\\
&&+\text{sign}({m_\ast}) \frac{T_L g_{-}(0)}{2 \pi v^2}\mathcal{D}_{T_R/T_L}\left(\frac{m_\ast v}{T_L}(\zeta+v)\right)+O(T_{L/R}^2)\,.
\eea
Note that if $k^+_\zeta\sim T_{R/L}$ this expression agrees with \eqref{eq:nonlinresm>0} and \eqref{eq:nonlinresm<0}, indeed in this case $\mathcal{D}_{T_R/T_L}\left(\frac{m_\ast v}{T_L}(\zeta+v)\right)$ is exponentially small. 


\begin{thebibliography}{99}
\bibitem{note1}
Note that the density of excitations with vanishing energy is large also for small temperatures. However, the interaction terms for such excitations are negligible with respect to the curvature~\cite{ImGl09-science,ImGl09,ImScGl-review}.
\end{thebibliography}
\end{document}